\documentclass[%
superscriptaddress,
preprint,
showpacs,
 amsmath,amssymb,
 aps,
pra,
]{revtex4-1}

\newcommand\beq{\begin{equation}}
\newcommand\eeq{\end{equation}}

\usepackage{graphicx}
\usepackage{dcolumn}
\usepackage{bm}

\begin{document}


\title{Boundary Effects of Weak Nonlocality in Multilayered Dielectric Metamaterials}

\author{Giuseppe Castaldi}
\affiliation{Fields \& Waves Lab, Department of Engineering, University of Sannio, I-82100 Benevento, Italy}
\author{Andrea Al\`u}
\affiliation{Photonics Initiative, Advanced Science Research Center, City University of New York, New York, NY 10031, USA}
\affiliation{Physics Program, Graduate Center, City University of New York, New York, NY 10026, USA}
\affiliation{Department of Electrical Engineering, City College of New York, New York, NY 10031, USA}
\author{Vincenzo Galdi}
\email{vgaldi@unisannio.it}
\affiliation{Fields \& Waves Lab, Department of Engineering, University of Sannio, I-82100 Benevento, Italy}

\date{\today}


\begin{abstract}
Nonlocal (spatial-dispersion) effects in multilayered metamaterials composed of periodic stacks of alternating, deeply subwavelength dielectric layers are known to be negligibly weak. Counterintuitively, under certain critical conditions, {\em weak nonlocality} may build up {\em strong boundary effects} that are not captured by conventional (local) effective-medium models based on simple mixing formulas. Here, we show that this phenomenon can be fruitfully studied and understood in terms of error propagation in the iterated maps of the trace and anti-trace of the optical transfer matrix of the multilayer. Our approach effectively parameterizes these peculiar effects via remarkably simple and insightful closed-form expressions, which enable direct identification of the critical parameters and regimes. We also show how these boundary effects can be captured by suitable nonlocal corrections.
\end{abstract}

\maketitle

\section{Introduction}
Away from the quantum regime, the macroscopic electromagnetic response of material media is typically modeled via constitutive relationships featuring a set of {\em intensive} properties, such as dielectric permittivity, electrical conductivity, and magnetic permeability \cite{Landau:1960eo}. Although these quantities clearly depend on the fine (atomic and molecular) structure of the medium, they do not punctually describe the strong field fluctuations on such fine scales, but only some suitably averaged behavior.

Besides being a cornerstone of the electrodynamics of continuous media \cite{Landau:1960eo}, the above {\em homogenization} concept is also heavily applied to the description of ``metamaterials'', i.e., composite materials made of subwavelength-sized (dielectric or metallic) inclusions in a host medium, which can be purposely designed in order to exhibit specific desired properties \cite{Capolino:2009vr,Cai:2010:om}.

In their arguably simplest conceivable form, homogenized models are based on mixing formulas (e.g., Maxwell-Garnett) that essentially depend on the inclusions' material constituents as well as their shapes, orientations and filling fractions, but not on their specific sizes and spatial arrangement \cite{Sihvola:1999em}. Such {\em effective medium theory} (EMT) is known to work especially well for dielectric structures featuring electrically small inclusions, whereas it may become significantly  inaccurate in the presence of metallic constituents and or inclusions with moderate electrical sizes. In these last cases, {\em nonlocal} corrections (in the form of spatial derivatives of the fields  or, equivalently, wavevector dependence in the constitutive relationships) are typically needed to account for the arising spatial dispersion (see, e.g., \cite{Silveirinha:2007mh,Elser:2007ne,Alu:2011:fp,Chebykin:2011ne,Chebykin:2012ne}).

Contrary to the conventional wisdom above, Herzig Sheinfux {\em et al.} \cite{Sheinfux:2014sm} recently pointed out a deceptively simple example of an {\em all-dielectric} multilayered metamaterial featuring {\em deeply subwavelength} layers that may exhibit peculiar boundary effects that are not captured by standard (local) EMT approaches. More specifically, under certain critical illumination conditions, the optical transmission (and reflection) of a finite-thickness slab of such metamaterial may differ substantially from the local EMT prediction, and may become ultrasensitive to the spatial order and/or size of the layers as well as to the addition or removal of a very thin layer (see also \cite{Andryieuski:2015ae}). 
These counterintuitive effects, experimentally demonstrated by Zhukovsky {\em et al.} \cite{Zhukovsky:2015ed}, have been attributed to the peculiar (interface-dominated) phase-accumulation mechanism in the structure \cite{Sheinfux:2014sm}, and have been shown to be potentially captured by suitable (possibly nonlocal and bianisotropic) corrections \cite{Popov:2016oa,Lei:2017rt}.

Besides the inherent academic interest in the homogenization aspects, the above ultrasensitivity phenomenon may open intriguing venues in applications such as optical sensing and switching \cite{Sheinfux:2014sm,Andryieuski:2015ae}. Moreover, similar mechanisms have also been shown to play a key role in establishing Anderson localization in disordered nanophotonic structures \cite{Sheinfux:2016cr,Sheinfux:2017oo}.

Against the above background, in this paper, we propose a simple and physically incisive modeling of the boundary effects that can be induced by weak nonlocality in multilayered dielectric metamaterials. Our approach, based on the trace and anti-trace map formalism \cite{Wang:2000ta}, directly relates the geometrical and constitutive parameters of interest to a set of meaningful observables through simple closed-form expressions. Besides elucidating the underlying mechanisms, this directly enables the identification of critical parameters and regimes, and naturally suggests possible nonlocal corrections capable of capturing the effects. 
 
Accordingly, the rest of the paper is structured as follows. In Sec. \ref{Sec:Formulation}, we introduce the problem geometry and background, and outline the mathematical formulation of our approach. In Sec. \ref{Sec:EMTB}, we apply the approach to the analytical  modeling of the boundary effects induced by weak nonlocality, and identify two distinct mechanisms. Moreover, we illustrate some representative results, and also explore possible nonlocal corrections.  Finally, in Sec. \ref{Sec:Conclusions}, we provide some concluding remarks and perspectives. Ancillary technical derivations are detailed in four Appendices.

\section{Background and Mathematical Formulation}
\label{Sec:Formulation}

\subsection{Geometry}
Referring to the schematic in Fig. \ref{Figure1}, as in \cite{Sheinfux:2014sm}, we consider a multilayered metamaterial consisting of a periodic arrangement of alternating dielectric layers (of infinite extent in the $x-y$ plane, and stacked along the $z$-direction), with relative permittivities $\varepsilon_a$ and $\varepsilon_b$ and thicknesses $d_a$ and $d_b$, respectively. More specifically, we assume a structure of finite thickness $L$ made of $n$ unit cells (bilayers) embedded in a homogeneous medium with relative permittivity $\varepsilon_e$, illuminated by a time-harmonic $[\exp(-i\omega t)]$, transversely electric (TE) polarized plane-wave with $y$-directed electric field, impinging from the exterior medium at an angle $\theta$ with respect to the $z$-axis. 
Accordingly, the relevant components of the impinging wavevector ${\bf k}_e$ can be expressed as
\beq
k_{ze}=k\sqrt{\varepsilon_e}\cos\theta,\quad k_x=k\sqrt{\varepsilon_e}\sin\theta,
\label{eq:ke}
\eeq
with $k=\omega/c$ denoting the vacuum wavenumber (and $c$ the corresponding wavespeed), and $k_x$ subject to momentum conservation \cite{Born:1999un}.
In what follows, we assume propagating fields in the exterior medium (i.e., real-valued incidence angles $\theta$),
lossless dielectric materials (i.e., real-valued, positive $\varepsilon_a$, $\varepsilon_b$, $\varepsilon_e$), and deeply subwavelength layer thicknesses (i.e., $d_a, d_b\ll \lambda$, with $\lambda=2\pi/k$ denoting the vacuum wavelength).

As it will be clearer hereafter, our assumption of considering the same exterior medium at the two ends of the metamaterial slab, though slightly less general than that in previous studies \cite{Sheinfux:2014sm,Popov:2016oa,Lei:2017rt}, allows an effective description of the underlying physical mechanisms in terms of a minimal number of parameters. Dealing with a more general scenario featuring two different exterior media only implies formal complications, but it does not add significant physical insight in the phenomenon.

\subsection{Local EMT Formulation}
Following the standard (local) EMT formulation \cite{Sihvola:1999em}, the above multilayered metamaterial can be modeled in terms of a homogenized, uniaxially anisotropic medium. For the assumed TE polarization, the relevant (in-plane) component of the resulting relative-permittivity tensor is given by \cite{Sihvola:1999em}
\beq
{\bar \varepsilon}_{\parallel}=
f_a\varepsilon_a+f_b\varepsilon_b,
\label{eq:EMT}
\eeq
with $f_a=d_a/d$ and $f_b=d_b/d=1-f_a$ denoting the filling fractions of the two material constituents, and $d=d_a+d_b$ the unit-cell thickness. Here and henceforth, the overline is used to tag EMT-based quantities.
In such effective medium, the longitudinal wavenumber is given by
\beq
{\bar k}_z=\sqrt{k^2 {\bar \varepsilon}_{\parallel}-k_x^2},
\label{eq:kzbar}
\eeq
with the (conserved) transverse component $k_x$ already defined in (\ref{eq:ke}). 

\subsection{Transfer-Matrix Formalism}
\begin{figure}
	\centering
	\includegraphics[width=12cm]{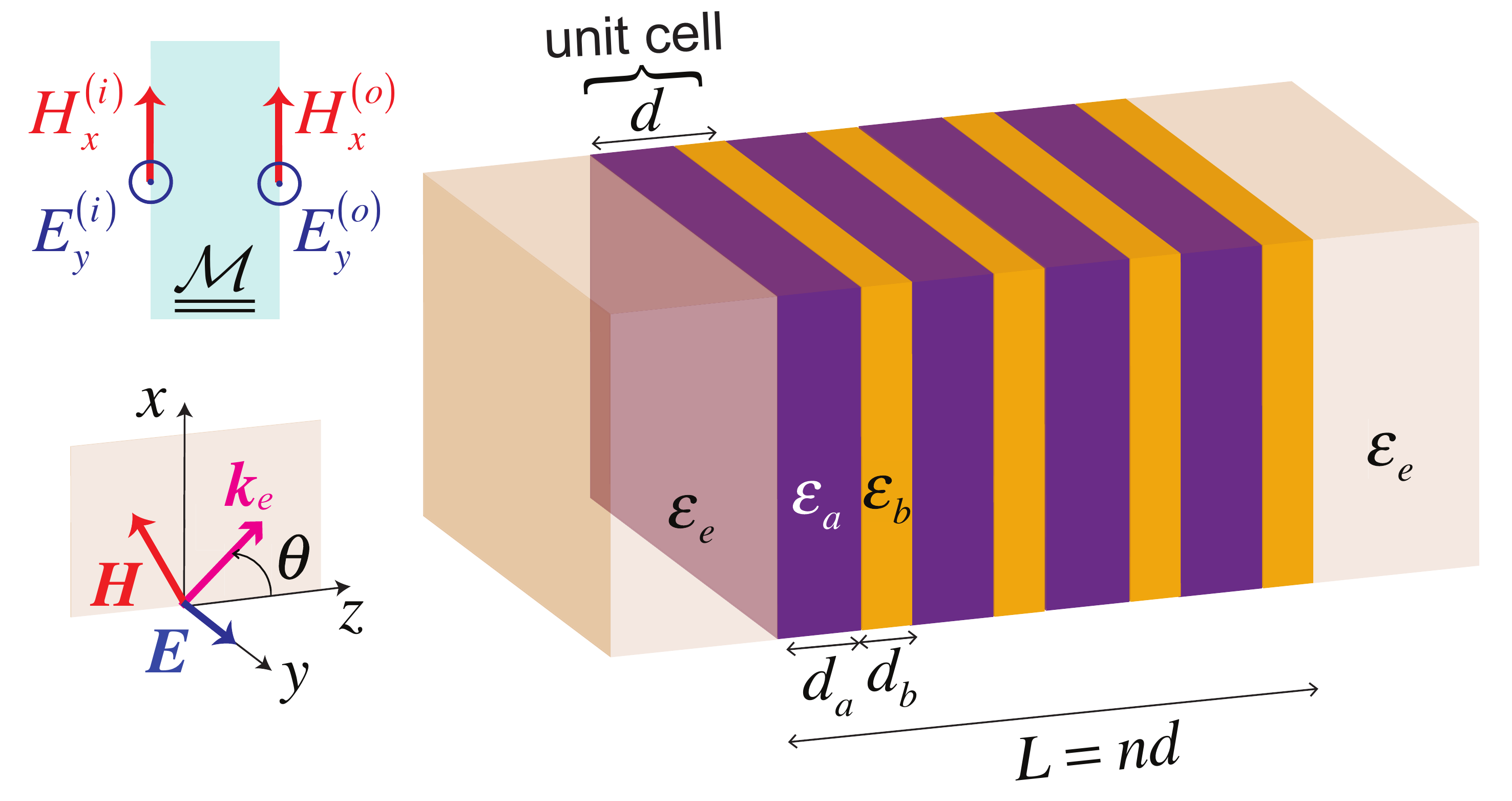}
	\caption{Problem geometry: A multilayered metamaterial composed of alternating dielectric layers, with relative permittivities $\varepsilon_a$ and $\varepsilon_b$, and thicknesses $d_a$ and $d_b$, respectively. The structure is assumed of infinite extent in the $x-y$ plane, and of finite thickness ($L=nd$) along the $z$-direction, and is embedded in a homogeneous medium with relative permittivity $\varepsilon_e$. The two insets on the left illustrate the transfer-matrix formalism [top, cf. (\ref{eq:TM})] and (bottom) the TE-polarized plane-wave illumination.}
	\label{Figure1}
\end{figure}

Our approach exploits as a rigorous reference solution the well-known transfer-matrix method \cite[Chap. 1]{Born:1999un}. As illustrated in the inset of Fig. \ref{Figure1}, for the assumed TE polarization, the tangential field components at two interfaces of a layer can be related via
\beq
\left[
\begin{array}{cc}
E_y^{(i)}\\
iZ_e H_x^{(i)}
\end{array}
\right]={\underline {\underline {\cal M}}}\cdot \left[
\begin{array}{cc}
E_y^{(o)}\\
iZ_e H_x^{(o)}
\end{array}
\right],
\label{eq:TM}
\eeq
where the superscripts $(i)$ and $(o)$ denote the input and output interfaces, respectively, ${\underline {\underline {\cal M}}}$ is a {\em unimodular}, dimensionless $2\times 2$ transfer matrix, and
\beq
Z_e=\frac{\omega \mu_0}{k_{ze}}
\eeq
represents the TE wave impedance in the exterior medium, with $\mu_0$ denoting the vacuum magnetic permeability, and $k_{ze}$ already defined in (\ref{eq:ke}). When multiple layers are cascaded, by iterating the above representation, the resulting transfer matrix can be obtained via chain product of the matrices representing the single layers \cite[Chap. 1]{Born:1999un}. Thus, for example, the transfer matrix pertaining to a unit cell as in Fig. \ref{Figure1} is given by
\beq
{\underline {\underline {\cal M}}}_{ab}={\underline {\underline {\cal M}}}_a\cdot
{\underline {\underline {\cal M}}}_b,
\label{eq:MMat}
\eeq
with the expressions of the matrices ${\underline {\underline {\cal M}}}_a$ and ${\underline {\underline {\cal M}}}_b$ explicitly given in Appendix \ref{App:Reference}.
Likewise, the transfer matrix pertaining to a multilayer composed of $n$ unit cells is straightforwardly obtained via a $n$-th power, viz.,
\beq
{\underline {\underline {\cal M}}}_n={\underline {\underline {\cal M}}}_{ab}^n.
\label{eq:Mn}
\eeq
From the transfer matrices in (\ref{eq:MMat}) and (\ref{eq:Mn}), the optical response of the multilayered metamaterial can be fully characterized, both 
in terms of Bloch-type dispersion relationship (in the infinite periodic limit)
and of transmission/reflection (for a finite number of layers) \cite[Chap. 1]{Born:1999un}.  

\subsection{Trace and Anti-trace Maps}
Interestingly, for the study of some fundamental aspects of the optical response of a multilayered metamaterial, two suitable combinations of the transfer-matrix elements, known as {\em trace} and {\em anti-trace}, are sufficient \cite{Wang:2000ta}. For a generic $2\times 2$ matrix
\beq
{\underline {\underline {\cal M}}}=\left[
\begin{array}{cc}
	{\cal M}_{11} & {\cal M}_{12}\\
	{\cal M}_{21} & {\cal M}_{22}
\end{array}
\right],
\eeq
they are defined as \cite{Lang:1987la}
\begin{subequations}
\begin{eqnarray}
\mbox{Tr}\left({\underline {\underline {\cal M}}}\right)&\equiv& {\cal M}_{11}+{\cal M}_{22},\\
\mbox{Atr}\left({\underline {\underline {\cal M}}}\right)&\equiv&{\cal M}_{21}-{\cal M}_{12},
\end{eqnarray}
\label{eq:TrAtr1}
\end{subequations}
respectively. For the periodic multilayer of interest here, by letting
\beq
\chi_n\equiv\mbox{Tr}\left({\underline {\underline {\cal M}}}_{ab}^n\right), \quad
\upsilon_n\equiv\mbox{Atr}\left({\underline {\underline {\cal M}}}_{ab}^n\right),
\label{eq:tat}
\eeq 
it can be shown \cite{Wang:2000ta} that the Bloch-type dispersion relationship (infinite number of layers) is given by 
\beq
\cos\left(k_{z}d\right)=\frac{\chi_1}{2}.
\label{eq:dispersion}
\eeq
Moreover, in the assumed scenario featuring the same exterior medium at the two ends (see Fig. \ref{Figure1}),
the transmission coefficient of a finite-size metamaterial slab comprising $n$ unit cells can be expressed as \cite{Wang:2000ta} (see also Appendix \ref{App:TraTr} for details)
\beq
\tau_n=\frac{2}{\chi_n+i\upsilon_n}.
\label{eq:taun}
\eeq
It is worth highlighting that, in view of the assumed lossless condition, the trace and anti-trace in (\ref{eq:tat}) are {\em real-valued} \cite{Wang:2000ta}.
By varying the number $n$ of unit cells, their evolution is governed by two iterated maps \cite{Wang:2000ta} (see also Appendix \ref{App:TraTr} for details),
\begin{subequations}
	\begin{eqnarray}
	\chi_n&=&\chi_1\chi_{n-1}-\chi_{n-2},\\
	\upsilon_n&=&\chi_1\upsilon_{n-1}-\upsilon_{n-2},\quad n\ge 2
	\end{eqnarray}
	\label{eq:chiupsilon}
\end{subequations}
which are coupled through the initial (unit-cell) trace $\chi_1$. Remarkably, the above iterated maps admit a closed-form analytical solution as \cite{Wang:2000ta} (see also Appendix \ref{App:TraTr} for details)
\begin{subequations}
\begin{eqnarray}
\chi_n&=&U_{n-1}\left(\frac{\chi_1}{2}\right)\chi_1-2 U_{n-2}\left(\frac{\chi_1}{2}\right)\nonumber\\
\label{eq:chi}
&=&U_{n}\left(\frac{\chi_1}{2}\right)-U_{n-2}\left(\frac{\chi_1}{2}\right)=2T_n\left(\frac{\chi_1}{2}\right),\\
\label{eq:upsilon}
\upsilon_n&=&U_{n-1}\left(\frac{\chi_1}{2}\right)\upsilon_1,
\end{eqnarray}
\label{eq:maps}
\end{subequations}
with $T_n$ and $U_n$ denoting Chebyshev polynomials of the first second kind, respectively \cite[Chap. 22]{Abramowitz:1965ho}.

We remark that, in spite of the different formalism adopted, the results in (\ref{eq:taun}) [with (\ref{eq:maps})] are exactly equivalent to those obtained via the conventional transfer-matrix method.

In what follows, we restrict our attention to the case $\chi_1\le2$ which, recalling the dispersion relationship in (\ref{eq:dispersion}), corresponds to the propagating condition (real-valued $k_z$) for the infinite multilayer. For finite-size structures, of direct interest in our study, by recalling (\ref{eq:chi}) and that $|T_n\left(\xi\right)|\le 1$ for $|\xi|\le 1$ \cite[Eq. 22.14.4]{Abramowitz:1965ho}, this also implies that 
\beq
|\chi_n|\le 2, \quad n\ge 1.
\label{eq:chibound}
\eeq

We stress that our assumption of identical exterior media at the two ends of the metamaterial does not imply a significant loss of generality in the description of the basic phenomena. If two different exterior media were assumed, as in \cite{Sheinfux:2014sm,Popov:2016oa,Lei:2017rt}, 
it would no longer be possible to express the transmission coefficient solely in terms of the trace and anti-trace. Aside from the arising formal complications, this would not hinder the applicability of our approach, since it is possible to derive iterated maps for all the terms of the transfer matrix \cite{Wang:2000ta}.

Moreover, it is worth highlighting that the above trace and anti-trace map formalism is not restricted to periodic multilayers, and can be extended to deal with rather general classes of {\em aperiodically ordered} structures generated by two-letter substitutional sequences (e.g., Fibonacci, Thue-Morse) \cite{Kolar:1990tm,Kolar:1990ma,Wang:2000ta,Savoia:2013on}, although the arising iterated maps do not generally admit analytical, closed-form solutions.

\section{Boundary Effects of Weak Nonlocality}
\label{Sec:EMTB}
\subsection{Background}
In \cite{Sheinfux:2014sm}, for the case $\varepsilon_e>{\bar \varepsilon}_{\parallel}$, it was observed that the agreement between the exact optical response of a finite-thickness multilayered metamaterial and the local EMT prediction would strongly deteriorate for incidence angles approaching the critical angle
\beq
\theta_c=\arcsin\left(\sqrt{\frac{{\bar \varepsilon}_{\parallel}}{\varepsilon_e}}\right)
\label{eq:thetac}
\eeq
which characterizes the total-internal-reflection condition between the exterior and effective media or, equivalently, the vanishing of the EMT-based longitudinal wavenumber ${\bar k}_z$ in (\ref{eq:kzbar}). Assuming $\theta\lesssim \theta_c$, in view of (\ref{eq:EMT}), this regime implies that the field is propagating in the higher-permittivity layers and is evanescent in the lower-permittivity ones. In \cite{Lei:2017rt}, an additional, independent mechanism was identified, leading to the breakdown of the standard EMT. This latter mechanism is not restricted to the critical-angle incidence above, but it is rather related to a phase-mismatch at the interface separating the last layer and the exterior medium, and becomes particularly significant for $\varepsilon_e={\bar \varepsilon}_{\parallel}$.

The breakdown of the local EMT model in the above scenarios is rather counterintuitive, as fully dielectric structures with deeply subwavelength inclusions are
known to exhibit {\em negligibly weak} nonlocal effects. In fact, such breakdown appears to be attributable to {\em boundary effects}, as it is only manifested in the transmission (and reflection) response of finite-size structures, whereas the {\em bulk} properties (dispersion relationship) are still accurately captured by the local EMT prediction \cite{Sheinfux:2014sm,Zhukovsky:2015ed}.

\subsection{Connection with Trace and Anti-trace Map Formalism}

To effectively model the above phenomena, it is expedient to interpret the EMT homogenized structure as a fictitious multilayer composed of homogeneous unit cells of thickness $d$ and relative permittivity ${\bar \varepsilon}_{\parallel}$ in (\ref{eq:EMT}). By comparing the traces and anti-traces of the transfer matrices pertaining to the actual and homogenized unit cells, we obtain (see Appendix \ref{App:Delta} for details)
\begin{subequations}
\begin{eqnarray}
\Delta\chi_1&=&\chi_1-{\bar \chi}_1\approx-\frac{\left(kd\right)^4\left(\varepsilon_a-\varepsilon_b\right)^2f_a^2f_b^2}{12},
\label{eq:Deltachi1}\\
\Delta\upsilon_1&=&\upsilon_1-{\bar \upsilon}_1\approx \frac{k\left(kd\right)^3\left(\varepsilon_a-\varepsilon_b\right)f_af_b
	\left[
	\left(\varepsilon_a+\varepsilon_b-2\varepsilon_e
	\right)f_b-\varepsilon_a+\varepsilon_e
	\right]}{6k_{ze}},
\label{eq:Deltaupsilon1}
\end{eqnarray}
\label{eq:Delta1}
\end{subequations}
where, following the previously introduced notation, EMT-based quantities are indicated by an overline. The errors in (\ref{eq:Delta1}) are manifestations of the inherent nonlocality of multilayered metamaterials. In structures such as hyperbolic metamaterials made of metallo-dielectric multilayers, these errors can be of sizable magnitude even for deeply subwavelength layers \cite{Elser:2007ne}, thereby leading to strong bulk effects, such as additional extraordinary waves \cite{Orlov:2011eo}. Conversely, for the case of {\em fully dielectric}, deeply subwavelength layers ($kd\ll 1$) of interest here, these errors are usually quite small, and are consequently expected to yield second-order effects. This is especially true for the trace error in (\ref{eq:Deltachi1}), whereas (\ref{eq:Deltaupsilon1}) indicates that the anti-trace error may become not so small for $k_{ze}\ll k$, i.e., for near-grazing incidence from the exterior medium. 

We recall, from (\ref{eq:dispersion}), that the dispersion relationship depends solely on the initial (unit-cell) trace $\chi_1$; this explains why, in the assumed conditions of deeply subwavelength dielectric layers, the EMT prediction of the {\em bulk} properties remains uniformly accurate even in the critical regimes observed in \cite{Sheinfux:2014sm,Zhukovsky:2015ed,Lei:2017rt}. This is not necessarily the case for the transmission coefficient of a {\em finite-thickness} structure [cf. (\ref{eq:taun})], which instead depends on {\em both} the trace and anti-trace iterated maps. Even in the case of 
{\em negligibly weak} nonlocality, resulting into very
accurate approximations at the unit-cell level, i.e., $|\Delta\chi_1|, |\Delta\upsilon_1|\ll 1$, there is no guarantee that the errors will remain negligibly small as the iterations proceed (i.e., as the number of unit cells increases). This is the key observation behind our approach, from which
it emerges that the boundary effects observed in \cite{Sheinfux:2014sm,Zhukovsky:2015ed,Lei:2017rt}
can be incisively interpreted and parameterized as an {\em error-propagation} problem in the trace and anti-trace iterated maps. In particular, we will show that the propagation effects of the initial (unit-cell) trace and anti-trace errors ($\Delta\chi_1$ and $\Delta\upsilon_1$, respectively) are directly associated with the two distinct mechanisms identified in \cite{Sheinfux:2014sm,Lei:2017rt}.

\subsection{Analytical Modeling}

We define the error maps
\begin{subequations}
\begin{eqnarray}
\Delta\chi_n&\equiv&\chi_n-{\bar \chi}_n,
\label{eq:Deltachin1}\\
\Delta\upsilon_n&\equiv&\upsilon_n-{\bar \upsilon}_n,
\label{eq:Deltaupsilonn1}\\
\Delta \tau_n&\equiv& \tau_n-{\bar \tau}_n,
\label{eq:Deltataun1}
\end{eqnarray}
\label{eq:errmap}
\end{subequations}
which describe the propagation of the initial errors in (\ref{eq:Delta1}). In particular, for a metamaterial composed of $n$ bilayers, $\Delta \tau_n$ quantifies the departure of the EMT-based approximation of the transmission coefficient from the exact (transfer-matrix-based) prediction.
By assuming $|\Delta\chi_1|, |\Delta\upsilon_1|\ll 1$, and exploiting the analytical solutions in (\ref{eq:maps}) (see Appendix \ref{App:Delta} for details), we obtain
\begin{subequations}
	\begin{eqnarray}
	\Delta\chi_n&\approx&X\sin\!\left(n\kappa\right)
	\sin\left(n \Omega\right),
	\label{eq:Deltachi}\\
	\Delta\upsilon_n&\approx&\Upsilon\cos\!\left(n\kappa\right)
	\sin\left(n \Omega\right)+{\cal O}\left(\Delta\chi_1,\Delta\upsilon_1\right),
	\label{eq:DeltaUpsilon}
	\end{eqnarray}
	\label{eq:EMTB}
\end{subequations}
where ${\cal O}$ denotes the Landau ``big-O'' symbol, and
\beq
X= -4,\quad
\Upsilon= 2\left(
\frac{{\bar k}_z}{k_{ze}}+\frac{k_{ze}}{{\bar k}_z}\right),
\label{eq:DeltaXY}
\eeq
\begin{subequations}
\begin{eqnarray}
\kappa&=&{\bar k}_z d+{\cal O}\left(\Delta\chi_1\right),
\label{eq:kf}\\
\Omega&=&-\frac{\Delta\chi_1}{2\sqrt{4-{\bar \chi}_1^2}}
+{\cal O}\left(\Delta\chi_1^2\right)
\approx\frac{\left(kd\right)^4\left(\varepsilon_a-\varepsilon_b\right)^2f_a^2f_b^2}{48 {\bar k}_z d}.
\label{eq:kappas}
\end{eqnarray}
\label{eq:kappa}
\end{subequations}

We observe from (\ref{eq:EMTB})--(\ref{eq:kappa}) that the leading terms in the error maps depend on the initial (unit-cell) trace error $\Delta\chi_1$, whereas the initial anti-trace error $\Delta\upsilon_1$ only affects
the higher-order correction ${\cal O}\left(\Delta\chi_1,\Delta\upsilon_1\right)$ in (\ref{eq:DeltaUpsilon}). 
 Throughout the paper, we will refer to these boundary effects as ``Type-I'' and ``Type-II'', respectively.

\subsection{Type-I Boundary Effects}
We start considering the Type-I boundary effects, i.e., neglecting the higher-order correction ${\cal O}\left(\Delta\chi_1,\Delta\upsilon_1\right)$ in (\ref{eq:DeltaUpsilon}), and focusing on the $\Delta\chi_1$-dependent leading terms. Under this assumption, we observe that the trace and anti-trace errors propagate with oscillatory laws characterized by two scales ($\kappa$ and $\Omega$, identical for both), and two different amplitudes.
Remarkably, both amplitudes {\em do not} depend on the initial (unit-cell) errors. More specifically, the trace error is always $\le 4$ in absolute value, irrespective of the incidence conditions and exterior medium, which is a direct consequence of the inherent trace bound in (\ref{eq:chibound}). Conversely, it is readily verified from (\ref{eq:DeltaXY}) that the amplitude $\Upsilon$ of the anti-trace error oscillations is always $\ge 4$, and critically depends on the incidence conditions and exterior medium. In particular, it can become {\em arbitrarily large} in the two (opposite) limits ${\bar k}_z/k_{ze}\ll 1$ and $k_{ze}/{\bar k}_z\ll 1$. The former limit corresponds to the critical-angle-incidence condition [cf. (\ref{eq:thetac})] explored in \cite{Sheinfux:2014sm}. The latter limit, instead, becomes relevant when $\varepsilon_e<{\bar \varepsilon}_{\parallel}$ and, to the best of our knowledge, has not been observed and explored in past studies.
The above observations imply that, in their evolutions, the trace and anti-trace error maps can exhibit peaks that are well beyond unity, which, via (\ref{eq:taun}), translate in significant departures of the optical response from its EMT-based prediction, i.e., transmission-coefficient errors $\Delta\tau_n$ on the scale of unity.

Looking at the two scales that characterize the error oscillations, we observe from (\ref{eq:kappa}) that $\kappa$ essentially accounts for the phase accumulation in the EMT-homogenized medium, with a small correction on the order of $\Delta\chi_1$, whereas $\Omega$ directly depends on $\Delta\chi_1$, with a higher-order correction on the order of $(\Delta\chi_1)^2$ (see Appendix \ref{App:Delta} for details).
  For typical parameter ranges of interest in this study, the two scales may be markedly different and, in particular, $\Omega\ll \kappa$. This yields a {\em slowly varying} envelope (ruled by $\Omega$, and identical for both the trace and anti-trace errors) modulated by {\em fast} oscillations (in quadrature, and ruled by $\kappa$). In this case, the slow scale $\Omega$ is particularly meaningful to understand and parameterize the Type-I boundary effects, and 
we can estimate from (\ref{eq:kappas}) the {\em critical size}, i.e., the number of unit cells (apart from periodicities)
\beq
n_p=\frac{\pi}{2\Omega}\approx \frac{24 \pi {\bar k}_zd}{\left(kd\right)^4\left(\varepsilon_a-\varepsilon_b\right)^2
f_a^2 f_b^2}
\label{eq:nn}
\eeq
around which the errors assume their peak values and, hence, the transmission-coefficient error (magnitude) $|\Delta\tau_n|$ is maximum. 
Likewise, it is apparent that for sizes $\approx 2n_p$ (plus periodicities) the errors tend to vanish.

We observe from (\ref{eq:DeltaXY}) that the anti-trace peak error $\Upsilon$ depends on the effective and exterior relative permittivities (${\bar \varepsilon}_{\parallel}$ and $\varepsilon_e$, respectively) as well as on the incidence angle. However, it {\em does not} depend on how the metamaterial is synthesized, i.e., the actual multilayer parameters ($\varepsilon_a$, $\varepsilon_b$, $d/\lambda$, $f_a$ and $f_b$), and hence the degree on nonlocality.
This latter, instead, plays a key role in establishing the critical size in (\ref{eq:nn}).
In other words, however weak the degree of nonlocality (i.e., however small $\Delta\chi_1$), once the desired effective and exterior permittivities and incidence conditions are set, a maximum potential strength of the Type-I boundary effects is inherently established, which can become significantly large in certain critical conditions. The degree of nonlocality only affects the spatial scale for which the effects are manifested.

As an illustrative example, similar to the study in \cite{Sheinfux:2014sm}, we consider a multilayered metamaterial with parameters $\varepsilon_a=1$, $\varepsilon_b=5$, $d_a=d_b=0.02\lambda$ (i.e., $f_a=f_b=0.5$, $d=\lambda/25$). We start considering a scenario featuring an exterior medium with $\varepsilon_e=4$ and an incidence angle $\theta=59^o$ close to the critical angle [$\theta_c=60^o$, from (\ref{eq:thetac})]. We observe that, for this parameter configuration, the initial (unit-cell) trace and anti-trace errors are both very small ($\Delta\chi_1=-3.32\times10^{-4}$ and $\Delta\upsilon_1=-5.14\times 10^{-3}$), and we therefore expect the leading terms in (\ref{eq:EMTB}) (i.e., Type-I effects) to be dominant. Accordingly, since ${\bar k}_z/k_{ze}=0.240$ (i.e., $\Upsilon\approx8.82$), we expect sensible departures from the EMT predictions. Moreover, since $\Omega=1.34\times 10^{-3}\ll \kappa=6.21\times10^{-2}$, we expect the fast/slow-scale interpretation to hold.

\begin{figure}
	\centering
	\includegraphics[width=10cm]{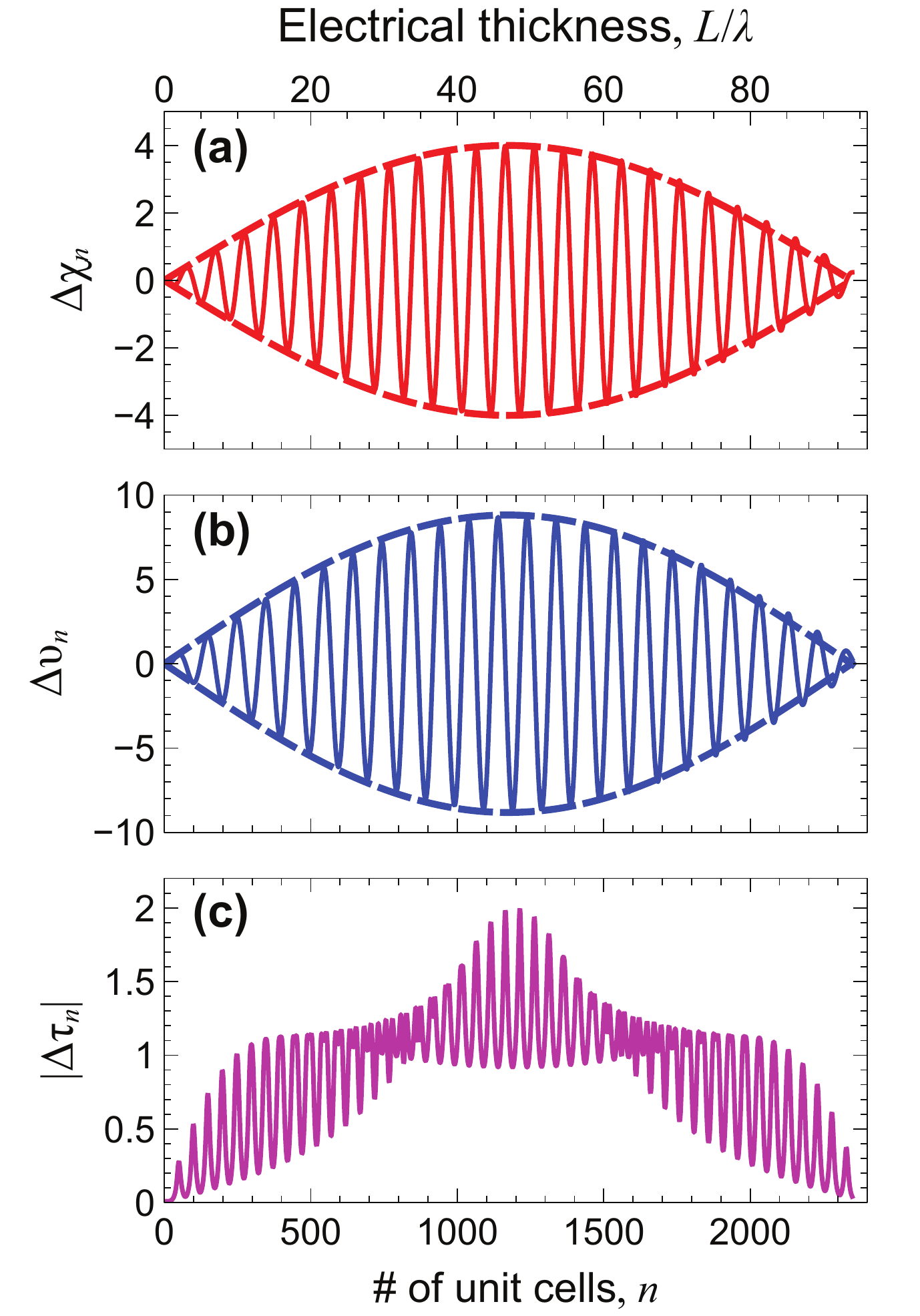}
	\caption{(a), (b) Trace and anti-trace error maps [(\ref{eq:Deltachin1}) and (\ref{eq:Deltaupsilonn1}), respectively], for a multilayer with $\varepsilon_a=1$, $\varepsilon_b=5$, $d_a=d_b=0.02\lambda$ (i.e., $d=\lambda/25$, $f_a=f_b=0.5$), an exterior medium with relative permittivity $\varepsilon_e=4$, and near-critical incidence angle $\theta=59^o$. The solid curves indicate the rigorous reference solutions, whereas the dashed curves indicate the slowly varying envelopes from the leading terms in (\ref{eq:EMTB}). (c) Corresponding transmission-coefficient difference map [(\ref{eq:Deltataun1})], computed via the rigorous reference solution. Results are shown as a function of the number of unit cells and corresponding electrical thickness $L/\lambda$ (shown on the upper horizontal axis).}
	\label{Figure2}
\end{figure}

Figure \ref{Figure2} illustrates the results pertaining to the error maps in (\ref{eq:errmap}). As expected, the error maps exhibit a two-scale oscillatory behavior, in very good agreement with the predictions in (\ref{eq:EMTB}). In particular, although it takes about 1200 unit cells (i.e., $L\approx 50\lambda$) to reach the peak error, sensible departures from the EMT predictions may be observed also for hundreds of unit cells (i.e., $L\approx 5\lambda$). As shown in Fig. \ref{Figure2}c, this corresponds to transmission-coefficient (magnitude) errors on the scale of unity.

\begin{figure}
	\centering
	\includegraphics[width=10cm]{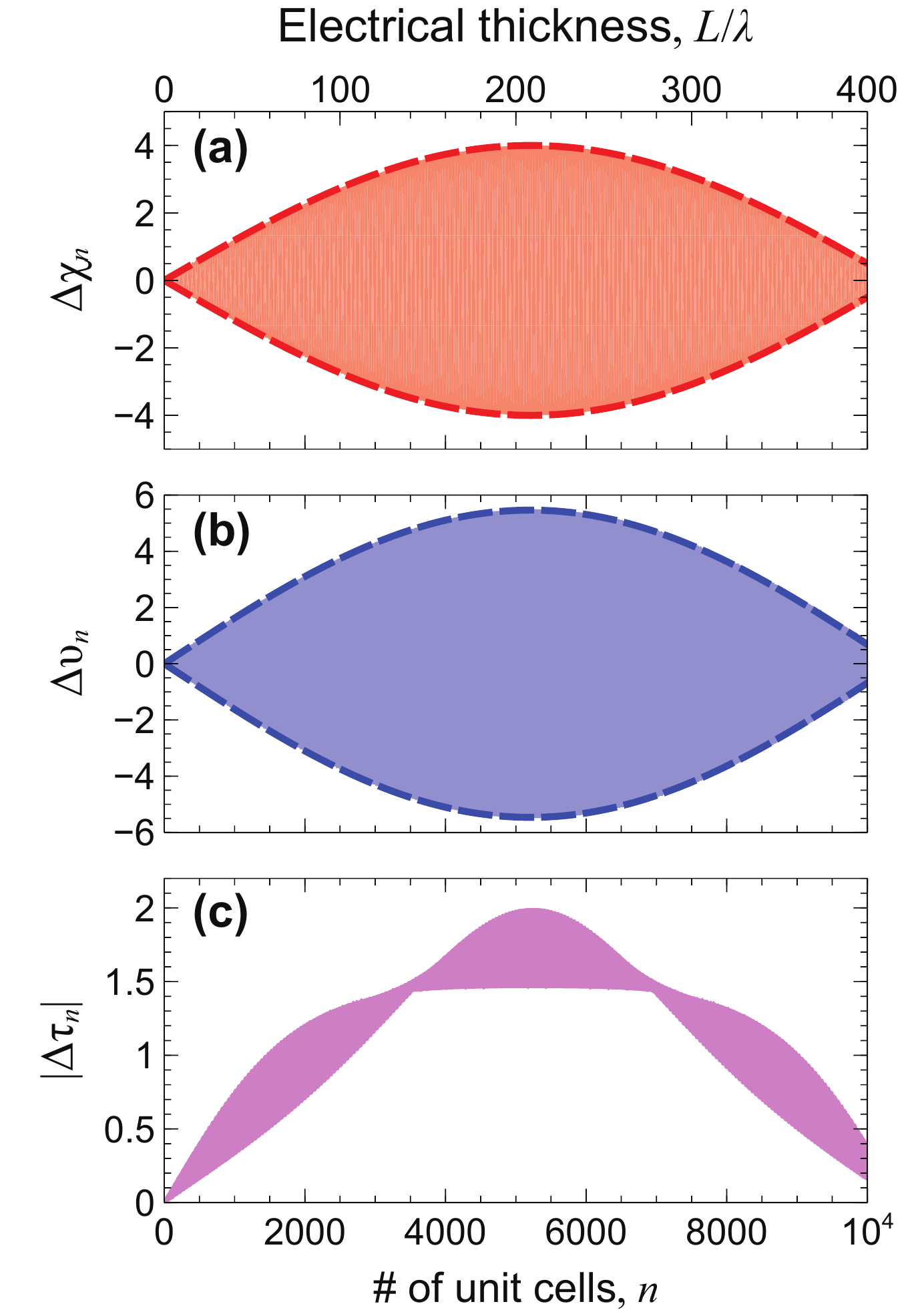}
	\caption{As in Fig. \ref{Figure2}, but with $\varepsilon_e=2$ and $\theta=70^o$. The shaded areas are representative of the very fast oscillations, which are not individually distinguishable on the scale of the plots.}
	\label{Figure3}
\end{figure}

For the same multilayer parameters as above, Fig. \ref{Figure3} shows some results pertaining to $\varepsilon_e=2$ and $\theta=70^o$, representative of the somehow opposite regime (${\bar k}_z/k_{ze}=2.30$), which was not considered in previous studies. We observe that the initial trace error $\Delta\chi_1$ is the same as the previous example (since it only depends on the multilayer parameters, which are not changed), whereas the anti-trace one is different ($\Delta\upsilon_1=-1.09\times10^{-2}$), but still sufficiently small for the Type-I effects to be dominant.
 Although total reflection is not possible in this scenario (since $\varepsilon_e<{\bar \varepsilon}_{\parallel}$), the field is still propagating in the higher-permittivity (``$b$''-type) layers and evanescent in the lower-permittivity (``$a$''-type) ones, and the Type-I boundary effects still become visible for sufficiently large sizes ($n_p=5275$). 
Note that, since the fast scale is now almost three-orders-of-magnitude larger that the slow one ($\kappa=937.7\Omega$), the fast oscillations are not individually distinguishable on the scale of the plots, and therefore we can only observe some shaded areas, whose envelopes are in very good agreement with our approximate modeling in (\ref{eq:EMTB}). Other than that, the same observations as for the previous example hold.

\subsection{Nonlocal Corrections}
\label{Sec:Mitigation}
The results above naturally suggest that the Type-I boundary effects can be captured by a suitable nonlocal correction providing a more accurate approximation of the initial (unit-cell) trace error $\Delta\chi_1$.

\begin{figure}
	\centering
	\includegraphics[width=10cm]{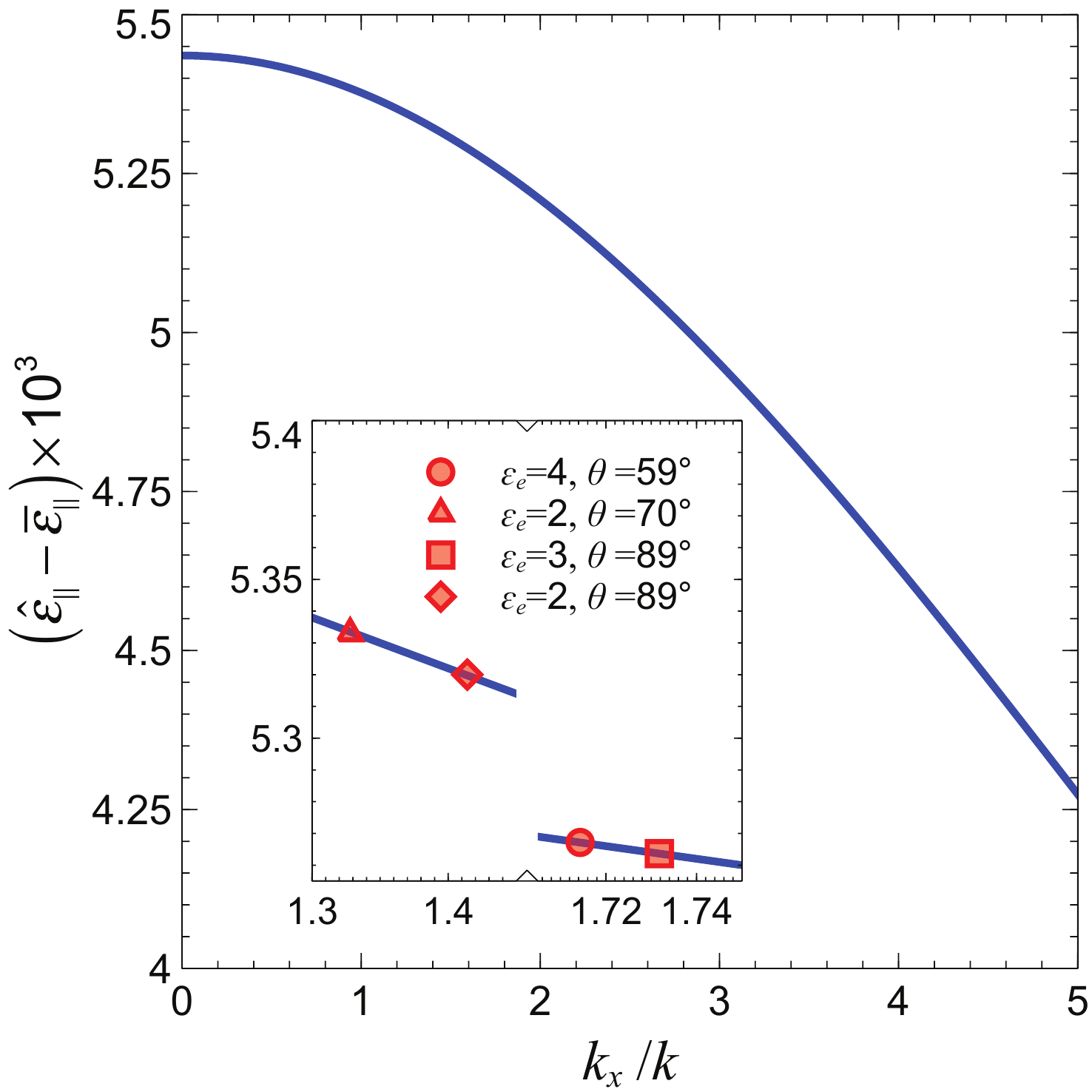}
	\caption{Nonlocal correction [cf. (\ref{eq:epsNL})] on the conventional EMT model in (\ref{eq:EMT}), as a function of the (normalized) transverse wavenumber, for a multilayer with $\varepsilon_a=1$, $\varepsilon_b=5$, $d_a=d_b=0.02\lambda$ (i.e., $d=\lambda/25$, $f_a=f_b=0.5$). Note the $10^3$ scale factor on the vertical axis. The magnified view in the inset shows the parameter ranges of main interest (note the broken horizontal axis), with the circle, triangle, square and diamond markers indicating the exterior-medium and incidence conditions considered in the various examples.}
	\label{Figure4}
\end{figure}

To this aim, along the lines of the approach proposed in \cite{Elser:2007ne}, we derive a {\em nonlocal} effective model (see Appendix \ref{App:NL} for details) in terms of a wavenumber-dependent relative permittivity
\begin{widetext}
\beq
{\hat \varepsilon}_{\parallel}\left(k_x\right) = \frac{{6 + k_x^2d^2 - \sqrt {36 + 12{d^2}\left(k_x^2 - {k^2}{\bar \varepsilon}_{\parallel}\right) + d^4\left(k_x^2 - \alpha_ak^2\right)\left(k_x^2 - \alpha_bk^2 \right)} }}{k^2d^2},
\label{eq:epsNL}
\eeq
\end{widetext}
with
\begin{subequations}
\begin{eqnarray}
\alpha_a&=& f_a^2\varepsilon_a+f_b^2\varepsilon_b+2f_af_b\varepsilon_a,\\
\alpha_b&=& f_a^2\varepsilon_a+f_b^2\varepsilon_b+2f_af_b\varepsilon_b.
\end{eqnarray}
\label{eq:alphas}
\end{subequations}
Here and henceforth, the caret is utilized to tag quantities related with the nonlocal effective model. It can be verified that, in the limit $kd\rightarrow 0$, the nonlocal model in (\ref{eq:epsNL}) consistently reduces to the conventional (local) EMT model in (\ref{eq:EMT}).
In fact, as illustrated in Fig. \ref{Figure4}, for the finite but small values of $kd$ of interest here, the actual nonlocal corrections are very small (on the third decimal figure), thereby confirming the anticipated weak character. As a consequence, strong-nonlocality-induced effects such as additional extraordinary waves should not be expected, and there is no need to enforce additional boundary conditions when solving the boundary-value problem.
 For our specific case, 
it can be shown (see Appendix \ref{App:NL} for details) that the initial (unit-cell) trace error arising from (\ref{eq:epsNL}) is given by
\beq
\Delta\chi_1=\chi_1-
{\hat \chi}_1
\sim {\cal O}\left[
\left(kd\right)^6\left( {{\varepsilon_a} - {\varepsilon_b}} \right)^2f_a^2f_b^2
\right].
\label{eq:hatDchi1}
\eeq
By comparison with the EMT-based counterpart in (\ref{eq:Deltachi1}), we observe a {\em much faster} decrease (sixth power, instead of fourth) of the error with the unit-cell electrical thickness $kd$ which, in the regime of interest, potentially translates [via (\ref{eq:kappas}) and (\ref{eq:nn})] into orders-of-magnitude increases of the critical size for the boundary effects to manifest.

\begin{figure}
	\centering
	\includegraphics[width=10cm]{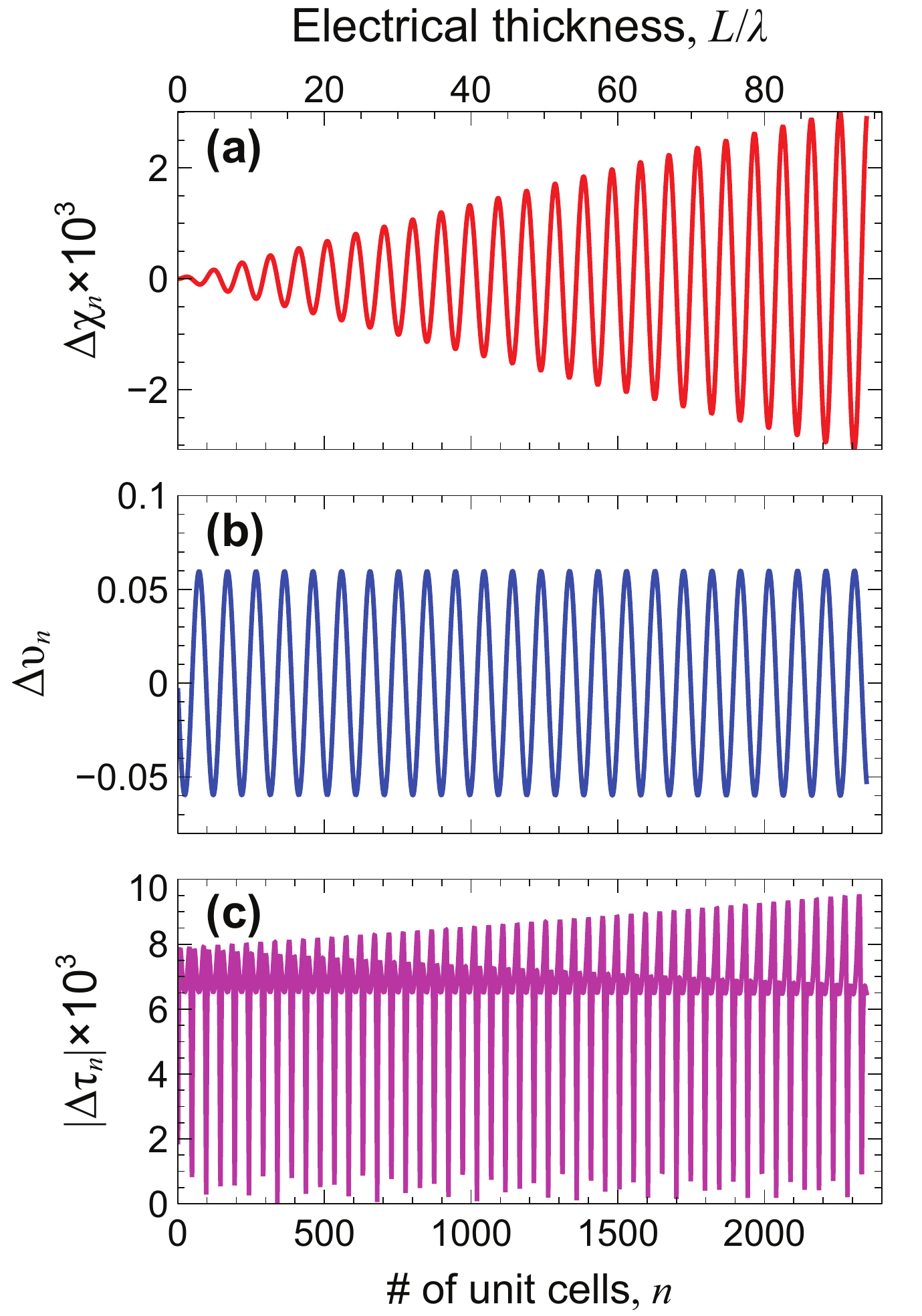}
	\caption{As in Fig. \ref{Figure2}, but considering the nonlocal effective model in (\ref{eq:epsNL}). Note the $10^3$ scale factors on the vertical axes in panels (a) and (c).}
	\label{Figure5}
\end{figure}

As an illustrative example, Fig. \ref{Figure5} shows the results corresponding to the configuration of Fig. \ref{Figure2}. For these parameters, the 
nonlocal effective model in (\ref{eq:epsNL}) yields a four-orders-of-magnitude smaller initial trace error ($\Delta\chi_1=8.55\times 10^{-8}$).
As a consequence, on the same spatial scale of Fig. \ref{Figure2}, both the trace (Fig. \ref{Figure5}a) and anti-trace (Fig. \ref{Figure5}b) errors are now strongly reduced, thereby yielding very small errors in the transmission coefficient (Fig. \ref{Figure5}c).

\begin{figure}
	\centering
	\includegraphics[width=10cm]{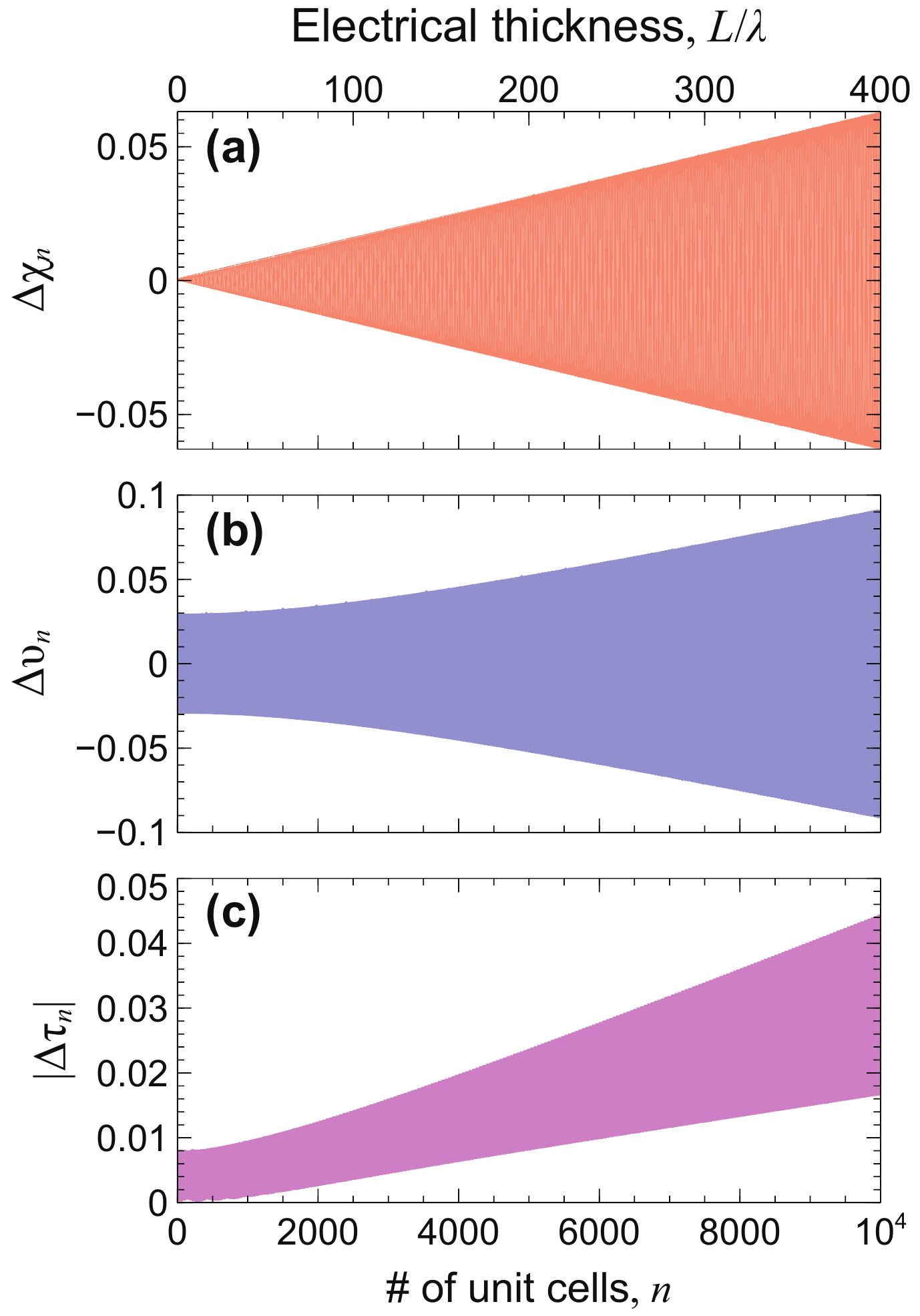}
	\caption{As in Fig. \ref{Figure3}, but considering the nonlocal effective model in (\ref{eq:epsNL}). Also in this case, the shaded areas are representative of the very fast oscillations, which are not individually distinguishable on the scale of the plots.}
	\label{Figure6}
\end{figure}

Qualitatively similar results can be observed in Fig. \ref{Figure6} for the parameter configuration of Fig. \ref{Figure3}. 

Clearly, higher-order nonlocal corrections could be derived \cite{Elser:2007ne}, which would further reduce the initial trace error $\Delta\chi_1$, and hence more accurately  capture the arising Type-I boundary effects. Alternatively, more sophisticated nonlocal effective models could be utilized \cite{Chebykin:2012ne}, which enforce the exact matching of the Bloch-type and effective dispersive relationships.

\subsection{Type-II Boundary Effects}
\label{Sec:ATI}
\begin{figure}
	\centering
	\includegraphics[width=10cm]{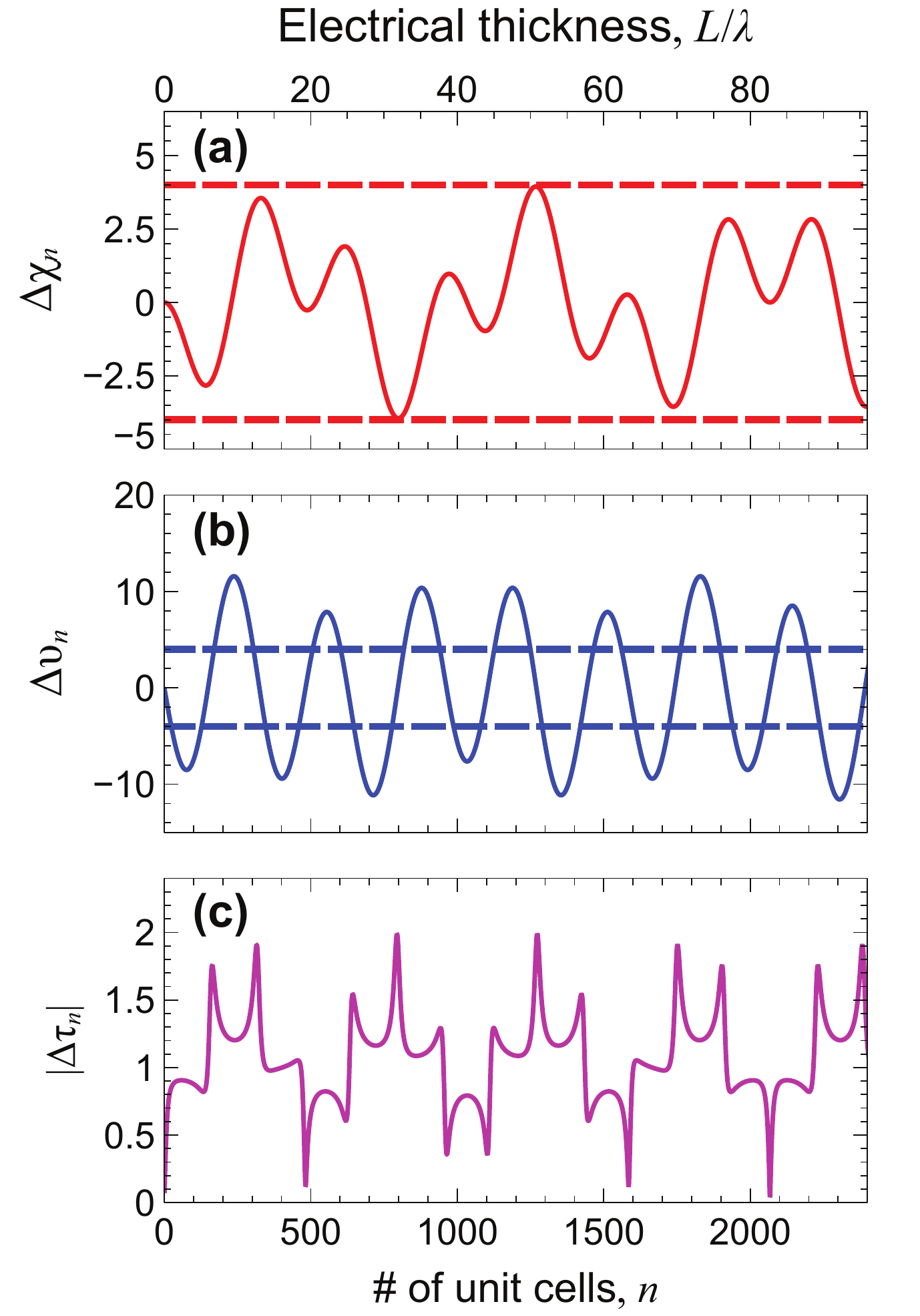}
	\caption{As in Fig. \ref{Figure2}, but with $\theta=89^o$ and $\varepsilon_e=3$. The dashed lines in panels (a) and (b) indicate the amplitude bounds [cf. (\ref{eq:DeltaXY})] predicted by the leading terms in (\ref{eq:EMTB}).}
	\label{Figure7}
\end{figure}

\begin{figure}
	\centering
	\includegraphics[width=10cm]{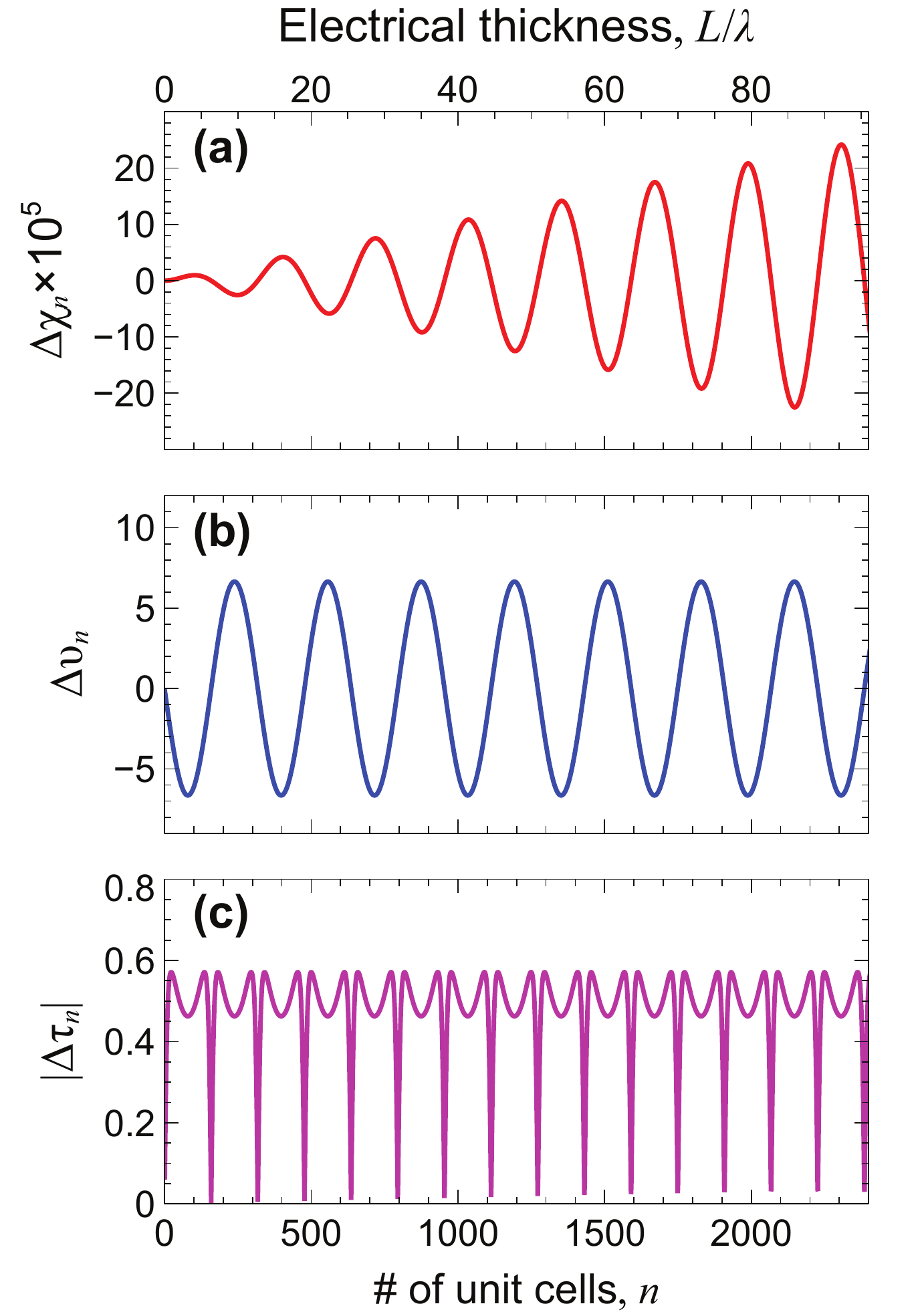}
	\caption{As in Fig. \ref{Figure7}, but considering the nonlocal effective model in (\ref{eq:epsNL}). Note the $10^5$ scale factor on the vertical axis in panel (a).}
	\label{Figure8}
\end{figure}

We now move on to consider the Type-II boundary effects, which become relevant in scenarios where 
the initial (unit-cell) anti-trace error $\Delta\upsilon_1$ and, hence,
the higher-order correction term in (\ref{eq:DeltaUpsilon}) are no longer negligibly small. 
From  (\ref{eq:Deltaupsilon1}), it is clear that this
 may happen, for instance, when $k_{ze}\ll k$, i.e., for near-grazing incidence from the exterior medium. This is exemplified in Figs. \ref{Figure7} and \ref{Figure8}, for a scenario with $\varepsilon_e={\bar\varepsilon}_{\parallel}=3$ and $\theta=89^o$, for which $\Delta\chi_1$ remains negligibly small, but $\Delta\upsilon_1=-0.18$ is no longer negligible. 
 
More specifically, Fig. \ref{Figure7} shows the error maps pertaining to the standard (local) EMT model. Unlike the previous examples (cf. Figs. \ref{Figure2} and \ref{Figure3}), we can no longer interpret the oscillatory behaviors of the trace and anti-trace errors in terms of fast oscillations and slow envelope, since the two scales are actually inverted ($\kappa<\Omega$) and rather close in value ($\kappa=7.60\times10^{-3}$, $\Omega=1.09\times10^{-2}$). Nevertheless, the trace error (Fig. \ref{Figure7}a) still satisfies the amplitude bound implied by (\ref{eq:Deltachi}). Conversely, we observe from Fig. \ref{Figure7}b that the anti-trace error may exceed (by more than a factor two) the amplitude bound predicted by $\Upsilon$ in (\ref{eq:DeltaXY}). As expected, these boundary effects are no longer accurately captured by the leading terms in (\ref{eq:EMTB}).

Figure \ref{Figure8} shows instead the corresponding results obtained by considering the nonlocal effective model in (\ref{eq:epsNL}). We observe that the trace error (Fig. \ref{Figure8}a) is now substantially reduced, but the anti-trace error (Fig. \ref{Figure8}b) still exhibits moderately large ($>5$) peak values. As a result, the transmission-coefficient error (Fig. \ref{Figure8}c) can still reach values $\sim 0.6$ in magnitude, even for relatively small sizes, thereby indicating that these boundary effects are not even captured very accurately by the nonlocal correction. This should not be surprising, as the nonlocal effective model in (\ref{eq:epsNL}) was derived with the aim of reducing the initial trace error ($\Delta\chi_1$) only, and it does not affect the initial anti-trace error $\Delta\upsilon_1$ responsible for the Type-II boundary effects.

\begin{figure}
	\centering
	\includegraphics[width=10cm]{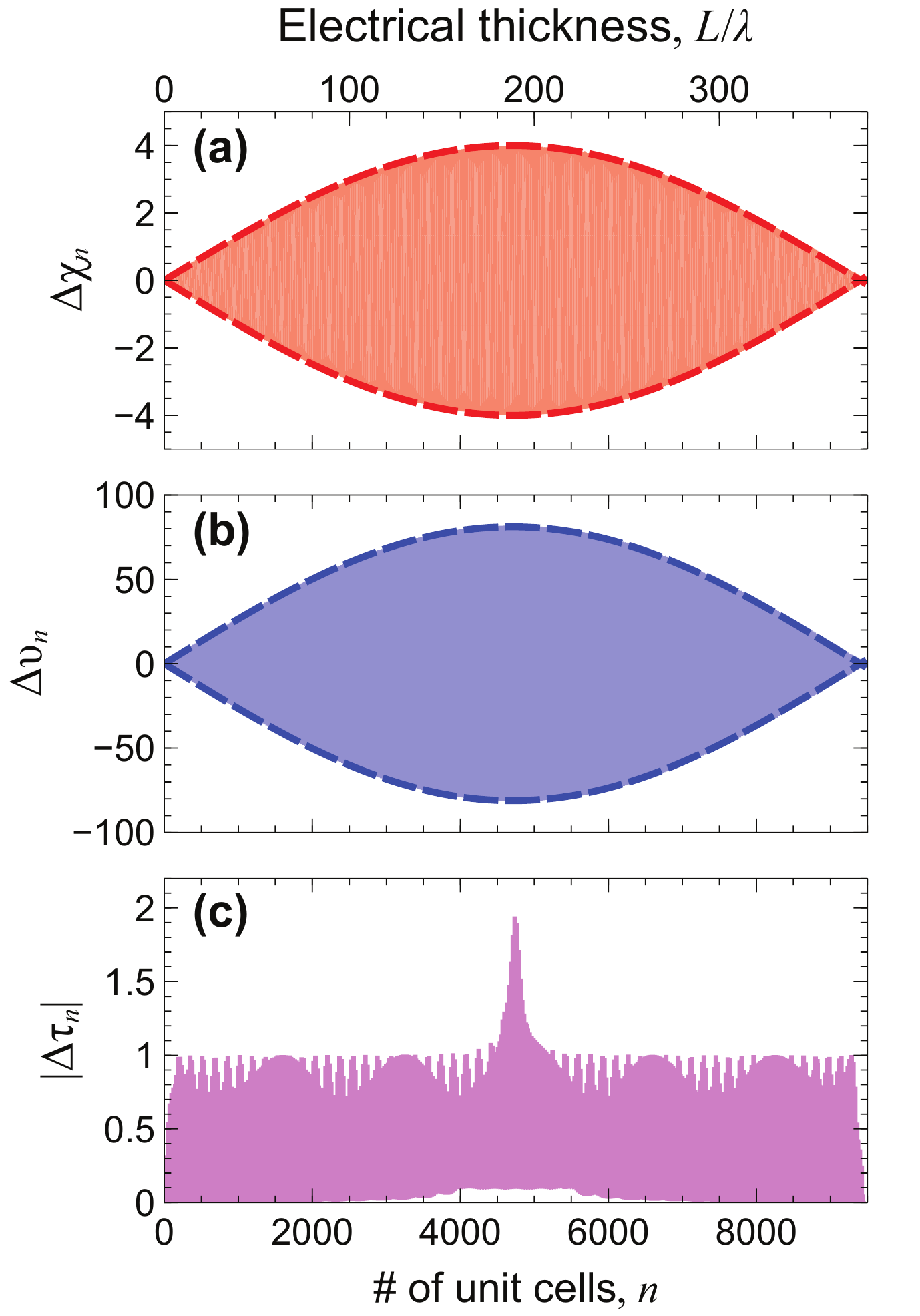}
	\caption{As in Fig. \ref{Figure3}, but with $\varepsilon_e=2$ and $\theta=89^o$. Also in this case, the shaded areas are representative of the very fast oscillations, which are not individually distinguishable on the scale of the plots.}
	\label{Figure9}
\end{figure}

\begin{figure}
	\centering
	\includegraphics[width=10cm]{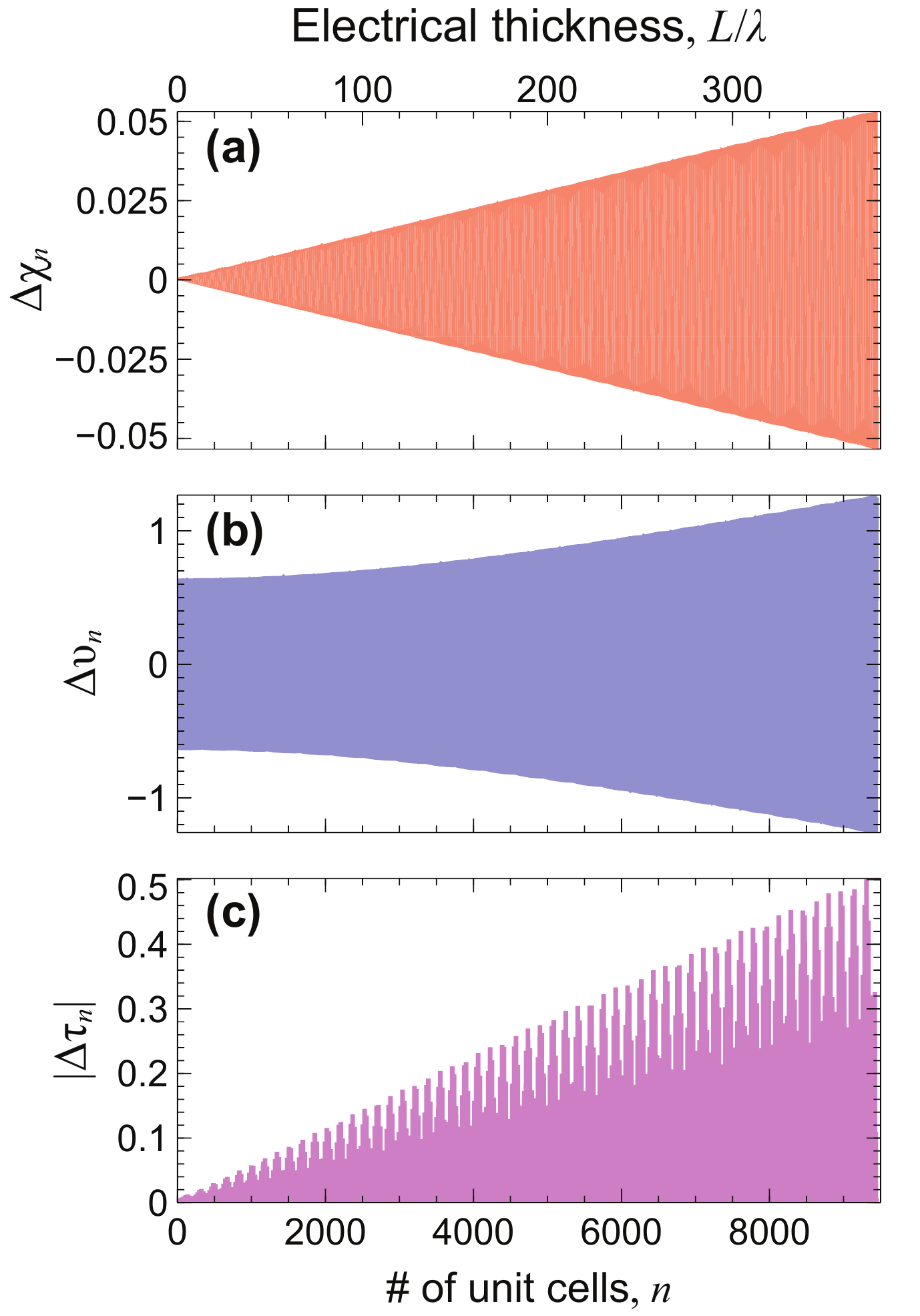}
	\caption{As in Fig. \ref{Figure9}, but considering the nonlocal effective model in (\ref{eq:epsNL}). Also in this case, the shaded areas are representative of the very fast oscillations, which are not individually distinguishable on the scale of the plots.}
	\label{Figure10}
\end{figure}

Another insightful example is illustrated in Figs. \ref{Figure9} and \ref{Figure10}, for a scenario with $\varepsilon_e=2$ and $\theta=89^o$. In this case, the initial anti-trace error $\Delta\upsilon_1=-0.21$, albeit not small in absolute terms, remains negligible by comparison with the quite large amplitude of the leading term ($\Upsilon=81$). In other words, in this regime, the Type-I boundary effects are still dominant.
As a result, we observe from Fig. \ref{Figure9} that the leading terms in (\ref{eq:EMTB}) still provide an accurate modeling. 
However, as clearly exemplified in Fig. \ref{Figure10}, when the nonlocal effective model is applied, and hence the Type-I effects are accurately captured, the residual Type-II effects become clearly visible.

The two examples above clearly indicate the different and independent 
nature of the Type-II boundary effects.
For the scenario $\varepsilon_e\ge {\bar \varepsilon}_{\parallel}$, a similar additional mechanism was identified in \cite{Lei:2017rt}. Such phenomenon,
not restricted to the critical-angle incidence and particularly significant for $\varepsilon_e={\bar \varepsilon}_{\parallel}$, was explained in terms of a phase-mismatch at the interface separating the last layer and the exterior medium. Remarkably, it was concluded that the effect could not be captured by local and nonlocal homogenization in terms of a {\em single} effective layer could not be applied, and the addition of an artificial matched layer was required. 

Our results indicate that such boundary effects may also become dominant in scenarios with $\varepsilon_e< {\bar \varepsilon}_{\parallel}$. Moreover, they
also indicate that a {\em single-parameter} nonlocal effective model is generally not sufficient to capture these effects. More 
complex (multiparameter) extensions are required in order to more accurately approximate {\em both} the initial (unit-cell) trace and anti-trace. 
However, since the anti-trace (unlike the trace) inherently depends on the exterior medium, the simplest conceivable extension, entailing the introduction of a (wavenumber-dependent) effective magnetic permeability would also inherit such dependence, thereby leading to a model inconsistency. A self-consistent model, with effective parameters independent of the exterior medium, should correctly approximate {\em all} the transfer-matrix terms. This would inevitably entail some magneto-electric coupling, along the lines of the approach in \cite{Popov:2016oa}. We regard this further development as beyond the scope of the present investigation.

\section{Conclusions and Outlook}
\label{Sec:Conclusions}
We have applied the trace and anti-trace map formalism to model the boundary effects induced by weak nonlocality in multilayered metamaterials made of periodic stacks of alternating, deeply subwavelength dielectric layers.

Our approach naturally identifies two distinct boundary effects, associated with different error propagation effects in the trace and anti-trace maps. 
Moreover, it leads to some analytical models that naturally highlight the critical parameters and conditions (including some not considered in previous studies), and provide very useful insights in the development of nonlocal corrections. 
Overall, we believe that our results nicely complement the previous studies on these phenomena \cite{Sheinfux:2014sm,Popov:2016oa,Lei:2017rt}, by offering a different perspective and paving the way to intriguing extensions.

For instance, our approach can be fruitfully extended to {\em aperiodically ordered} multilayers \cite{Kolar:1990tm,Kolar:1990ma,Wang:2000ta,Savoia:2013on}, in order to explore the effects induced by the spatial arrangement. Interestingly, due to the more complex character of the arising maps, the trace bound in (\ref{eq:chibound}) would not necessarily hold in these scenarios, thereby giving rise to richer dynamics involving transitions between propagating and evanescent regimes. Within a related framework, also of great interest are possible applications to the study of Anderson localization effects in random multilayers, along the lines of \cite{Sheinfux:2016cr,Sheinfux:2017oo}.

Finally, also worth of mention are possible extensions to non-Hermitian scenarios featuring balanced loss and gain, which are experiencing a surging interest in optics and photonics (see, e.g., \cite{Feng:2017ww} for a recent review). Within this framework, we observe that under the special conditions of {\em parity-time} symmetry \cite{Bender:1998rs} for the unit cell ($\varepsilon_a=\varepsilon_b^*$, $d_a=d_b$), the trace and anti-trace maps remain {\em real-valued}, thereby greatly simplifying the study and interpretation of the phenomena. Moreover, since the conventional EMT model of a parity-time-symmetric multilayer would yield a lossless (and gainless) effective medium, the phenomena studied here are expected to be closely related to distinctive effects that can occur in non-Hermitian systems, such as the emergence of {\em exceptional points} \cite{Heiss:2012th}.

\appendix

\section{Details on Transfer-Matrix Formalism}
\label{App:Reference}
Following \cite[Chap. 1]{Born:1999un}, the transfer matrices pertaining to the single layers can be expressed as
\begin{subequations}
\begin{eqnarray}
{\underline {\underline {\cal M}}}_{a}&=&
\left[
\begin{array}{cc}
	\cos\left(k_{za}f_ad\right) & \displaystyle{\frac{k_{ze}}{k_{za}}}\sin\left(k_{za}f_ad\right)\\
	-\displaystyle{\frac{k_{za}}{k_{ze}}}\sin\left(k_{za}f_ad\right)	 & \cos\left(k_{za}f_ad\right)
\end{array}
\right],\\
{\underline {\underline {\cal M}}}_{b}&=&
\left[
\begin{array}{cc}
	\cos\left(k_{zb}f_bd\right) & \displaystyle{\frac{k_{ze}}{k_{zb}}}\sin\left(k_{zb}f_bd\right)\\
	-\displaystyle{\frac{k_{zb}}{k_{ze}}}\sin\left(k_{zb}f_bd\right)	 & \cos\left(k_{zb}f_bd\right)
\end{array}
\right],
\end{eqnarray}
\end{subequations}
where $k_{za}=\sqrt{k^2\varepsilon_a-k_x^2}$ and $k_{zb}=\sqrt{k^2\varepsilon_b-k_x^2}$ denote the longitudinal wavenumbers in the two corresponding media, and all other symbols are already defined in the main text. By multiplying the two matrices above, the unit-cell trace and anti-trace are straightforwardly obtained as [cf. (\ref{eq:tat}) with $n=1$]
\begin{subequations}
\begin{eqnarray}
\chi_1&=&2 \cos\left(k_{za}f_ad\right)\cos\left(k_{zb}f_bd\right)
-
\left(
\frac{k_{za}}{k_{zb}}+\frac{k_{zb}}{k_{za}}\right)
\sin\left(k_{za}f_ad\right)
\sin\left(k_{zb}f_bd\right),
\label{eq:cchi1}
\\
\upsilon_1&=&-\left(
\frac{k_{za}}{k_{z}}+\frac{k_{z}}{k_{za}}\right)\sin\left(k_{za}f_ad\right)\cos\left(k_{zb}f_bd\right)\nonumber\\
&-&\left(\frac{k_{zb}}{k_{z}}+\frac{k_{z}}{k_{zb}}\right)\cos\left(k_{za}f_ad\right)\sin\left(k_{zb}f_bd\right).
\label{eq:uupsilon1}
\end{eqnarray}
\end{subequations}

\section{Details on Trace and Anti-trace Maps}
\label{App:TraTr}
Assuming a generic transfer matrix ${\underline {\underline {\cal M}}}$ (see the inset in Fig. \ref{Figure1}), and unit-amplitude, TE-polarized plane-wave incidence, the input and output electric fields can be written as 
\begin{subequations}
	\begin{eqnarray}
	E_y^{(i)}&=&\exp\left(i k_x x\right)\left[
	\exp\left(
	ik_{ze}z
	\right)+\Gamma \exp\left(
	-ik_{ze}z
	\right)\right],\\
	E_y^{(o)}&=&\tau
	\exp\left\{
	i\left[k_x x+
	k_{ze}\left(z-L\right)\right]\right\},
	\end{eqnarray}
\end{subequations}
where $\Gamma$ and $\tau$ denote the reflection and transmission coefficients, respectively. By computing the corresponding magnetic-field tangential components via the relevant Maxwell's curl equation, and substituting in (\ref{eq:TM}), we obtain the linear system
\beq
\left[
\begin{array}{cc}
	1+\Gamma\\
	-i\left(1-\Gamma\right)
\end{array}
\right]={\underline {\underline {\cal M}}}\cdot \left[
\begin{array}{cc}
	\tau\\
	-i\tau
\end{array}
\right],
\label{eq:TM1}
\eeq
which, solved with respect to $\tau$, yields
\beq
\tau=\frac{2}{{\cal M}_{11}+{\cal M}_{22}+i\left({\cal M}_{21}-{\cal M}_{12}\right)}. 
\label{eq:tau}
\eeq
Equation (\ref{eq:taun}) directly follows from (\ref{eq:tau}) by particularizing ${\underline {\underline {\cal M}}}={\underline {\underline {\cal M}}}_{ab}^n$, and recalling the trace and anti-trace definitions in (\ref{eq:TrAtr1}).

As a consequence of the Cayley-Hamilton theorem \cite{Lang:1987la}, the square of a $2\times 2$ unimodular matrix can be expressed as \cite{Kolar:1990ma}
\beq
{\underline {\underline {\cal M}}}^2=\mbox{Tr}\left({\underline {\underline {\cal M}}}\right) {\underline {\underline {\cal M}}}-{\underline {\underline {\cal I}}},
\label{eq:CH}
\eeq
with ${\underline {\underline {\cal I}}}$ denoting the $2\times 2$ identity matrix. The trace and anti-trace maps in (\ref{eq:chiupsilon}) are obtained from (\ref{eq:CH}), particularized for ${\underline {\underline {\cal M}}}={\underline {\underline {\cal M}}}_{ab}$, by multiplying both sides by ${\underline {\underline {\cal M}}}^{n-2}$ (with $n\ge 2$), and calculating the trace and anti-trace, respectively.

Via recursive application of (\ref{eq:CH}), we also obtain \cite{Kolar:1990ma}
\beq
{\underline {\underline {\cal M}}}^n=U_{n-1}\left[\frac{1}{2}\mbox{Tr}\left({\underline {\underline {\cal M}}}\right)\right] {\underline {\underline {\cal M}}}-U_{n-2}\left[\frac{1}{2}\mbox{Tr}\left({\underline {\underline {\cal M}}}\right)\right]{\underline {\underline {\cal I}}},
\label{eq:CH1}
\eeq
from which the analytical solutions in (\ref{eq:maps}) directly follow by particularizing ${\underline {\underline {\cal M}}}={\underline {\underline {\cal M}}}_{ab}$, and calculating the trace and anti-trace.

\section{Derivation of Eqs. (\ref{eq:Delta1}) and (\ref{eq:EMTB})}
\label{App:Delta}
First, we derive the approximations in (\ref{eq:Delta1}) for the initial (unit-cell) trace and anti-trace errors. By expanding (\ref{eq:cchi1}) in McLaurin series with respect to $d$ (up to the fourth order), we obtain
\beq
\chi_1\approx
2-\left({\bar k}_zd\right)^2+\frac{d^4}{12}
\left[
k_x^2-k^2\left(
f_a^2 \varepsilon_a+f_b^2\varepsilon_b+2f_af_b\varepsilon_a
\right)
\right]
\left[
k_x^2-k^2\left(
f_a^2 \varepsilon_a+f_b^2\varepsilon_b+2f_af_b\varepsilon_b
\right)
\right].
\label{eq:Chi1}
\eeq
Recalling the transfer matrix of the EMT-homogenized medium,
\beq
{\bar {\underline {\underline {\cal M}}}}=
\left[
\begin{array}{cc}
	\cos\left({\bar k}_zd\right) & \displaystyle{\frac{k_{ze}}{{\bar k}_z}}\sin\left({\bar k}_zd\right)\\
	-\displaystyle{\frac{{\bar k}_z}{k_{ze}}}\sin\left({\bar k}_zd\right)	 & \cos\left({\bar k}_zd\right)
\end{array}
\right],
\label{eq:Mbar}
\eeq
we obtain likewise (for ${\bar k}_zd \ll1$)
\beq
{\bar \chi}_1=2\cos\left({\bar k}_zd\right)\approx2-\left({\bar k}_zd\right)^2+\frac{\left({\bar k}_zd\right)^4}{12}.
\label{eq:Chi1b}
\eeq
The approximation in (\ref{eq:Deltachi1}) follows by subtracting (\ref{eq:Chi1}) and (\ref{eq:Chi1b}), and recalling (\ref{eq:kzbar}).

In a similar fashion, via McLaurin expansions of $\upsilon_1$ and ${\bar \upsilon}_1$ (up to the third order in $d$), we obtain 
\begin{eqnarray}
\upsilon_1&\approx& 
\frac{d}{6k_{ze}}
\left\{
f_a\left[
k^2\left(
\varepsilon_e+\varepsilon_a
\right)-2k_x^2
\right]
\left(
k_{za}^2d^2 f_a^2-6\right)
\right.\nonumber\\
&+&3f_b
\left[
k^2\left(
\varepsilon_e+\varepsilon_b
\right)-2k_x^2
\right]\left(
k_{za}^2d^2 f_a^2-2\right)\nonumber\\
&+&3k_{zb}^2d^2f_af_b^2
\left[
k^2\left(
\varepsilon_e+\varepsilon_a
\right)-2k_x^2
\right]\nonumber\\
&+&k_{zb}^2d^2f_b^3\left[
\left.
k^2\left(
\varepsilon_e+\varepsilon_b
\right)-2k_x^2
\right]\right\},
\end{eqnarray}
\beq
{\bar \upsilon}_1=-\left(
\frac{{\bar k}_z}{k_{ze}}+
\frac{k_{ze}}{{\bar k}_z}
\right)
\sin\left({\bar k}_zd\right)\approx \frac
{d\left[
	\left(
	{\bar \varepsilon}_{\parallel}+\varepsilon_e
	\right)k^2-2k_x^2
	\right]
\left(
{\bar k}_z^2d^2-6
\right)
}
{6k_{ze}},
\label{eq:bups1}
\eeq
which, after some algebra, yield the approximation in (\ref{eq:Deltaupsilon1}).

From (\ref{eq:Deltachin1}), by recalling the trigonometric form of the Chebyshev polynomials of the first kind \cite[Eq. 22.3.15]{Abramowitz:1965ho}, we obtain 
\begin{widetext}
\begin{eqnarray}
\Delta \chi_n&=& 2\left\{
\cos \left[n\arccos\left(\frac{{\bar \chi_1}\!+\!\Delta\chi_1}{2}\right)\right]
\!-\! \cos \left[n\arccos\left(\frac{{\bar \chi_1}}{2}\right)\right]\right\}
\nonumber\\
&=&  \!- 4\sin \left\{\frac{n}{2}
\left[\arccos\left(\frac{{\bar \chi_1}\!+\!\Delta\chi_1}{2}\right)
\!+\!\arccos\left(\frac{{\bar \chi_1}}{2}\right)\right]
\right\}\nonumber\\
&\times&\sin \left\{\frac{n}{2}
\left[\arccos\left(\frac{{\bar \chi_1}\!+\!\Delta\chi_1}{2}\right)
\!-\!\arccos\left(\frac{{\bar \chi_1}}{2}\right)\right]
\right\},
\label{eq:DDchi1}
\end{eqnarray}
\end{widetext}
where a
sum-to-product identity \cite[Eq. 4.3.37]{Abramowitz:1965ho} was exploited in the second equality.
Next, under the assumption $\Delta\chi_1\ll 1$, we apply the following approximations:
\begin{subequations}
\begin{eqnarray}
\arccos\left(\frac{{\bar \chi_1}\!+\!\Delta\chi_1}{2}\right)
+\arccos\left(\frac{{\bar \chi_1}}{2}\right)&\approx& 2 \arccos\left(\frac{{\bar\chi}_1}{2}\right)-\frac{\Delta\chi_1}{\sqrt{4-\chi_1^2}},\\
\arccos\left(\frac{{\bar \chi_1}\!+\!\Delta\chi_1}{2}\right)
-\arccos\left(\frac{{\bar \chi_1}}{2}\right)&\approx&-\frac{\Delta\chi_1}{\sqrt{4-\chi_1^2}}-\frac{\chi_1\left(\Delta\chi_1\right)^2}{2\sqrt{\left(4-\chi_1^2\right)^3}},
\end{eqnarray}
\label{eq:acos}
\end{subequations}
which, substituted in (\ref{eq:DDchi1}), yield
the parameterization in (\ref{eq:Deltachi}), with the expression of $\kappa$ and $\Omega$ in (\ref{eq:kappa}) obtained by recalling [from (\ref{eq:Deltachi1})] that
\beq
{\bar \chi}_1=2\cos\left({\bar k}_zd\right),
\eeq
and that, for ${\bar k}_zd\ll 1$,
\beq
{\sqrt{4-{\bar \chi}_1^2}}=2\sin\left({\bar k}_zd\right)\approx 2{\bar k}_zd.
\label{eq:sqrt}
\eeq

Moving on to the anti-trace error map in (\ref{eq:Deltaupsilonn1}), by recalling the trigonometric form of the Chebyshev polynomials of the second kind
\cite[Eq. 22.3.16]{Abramowitz:1965ho}, we obtain
\begin{widetext}
	\beq
	\Delta \upsilon_n= \frac{2\left({\bar \upsilon}_1+\Delta\upsilon_1\right)\sin \left[n\arccos\left(
		\displaystyle{\frac{{\bar \chi_1}\!+\!\Delta\chi_1}{2}}\right)\right]}{\sqrt{4-\left({\bar \chi}_1+\Delta\chi_1\right)^2}}
	\!-\! \frac{2{\bar \upsilon}_1\sin \left[n\arccos\left(
		\displaystyle{\frac{{\bar \chi_1}}{2}}\right)\right]}{\sqrt{4-{\bar \chi}_1^2}}.
	\label{eq:DDU}
	\eeq
\end{widetext}
Then, in the limit $|\Delta\chi_1|, |\Delta\upsilon_1|\ll1$, we exploit the following approximation
\beq
\frac{2\left({\bar \upsilon}_1+\Delta\upsilon_1\right)}{\sqrt{4-\left({\bar \chi}_1+\Delta\chi_1\right)^2}}
\approx
\frac{2{\bar \upsilon}_1}{\sqrt{4-{\bar \chi}_1^2}}+
\frac{2\Delta{\bar \upsilon}_1}{\sqrt{4-{\bar \chi}_1^2}}+
\frac{2{\bar \chi}_1{\bar \upsilon}_1\Delta\chi_1}{\sqrt{\left(4-{\bar \chi}_1^2\right)^3}},
\eeq
which, substituted in (\ref{eq:DDU}), yields the parameterization in (\ref{eq:DeltaUpsilon}),
by applying the same approximations as in (\ref{eq:acos}), and recalling that [from (\ref{eq:bups1}) and (\ref{eq:sqrt})]
\beq
\frac{2{\bar \upsilon}_1}{{\sqrt{4-{\bar \chi}_1^2}}}=-\left(
\frac{{\bar k}_z}{k_{ze}}+
\frac{k_{ze}}{{\bar k}_z}
\right).
\eeq

\section{Derivation of Eq. (\ref{eq:epsNL})}
\label{App:NL}
Our nonlocal homogenization strategy is inspired by the approach put forward in \cite{Elser:2007ne}. First,
similar to (\ref{eq:Chi1b}), we expand the trace of the transfer matrix of the nonlocally homogenized unit cell as
\beq
{\hat \chi}_1=2\cos\left({\hat k}_zd\right)\approx2-\left({\hat k}_zd\right)^2+\frac{\left({\hat k}_zd\right)^4}{12},
\label{eq:hatchi1}
\eeq
where, 
\beq
{\hat k}_z=\sqrt{k^2{\hat \varepsilon}_{\parallel}\left(k_x\right)-k_x^2},
\eeq
and, following the notation introduced in the main text, the caret denotes quantities based on nonlocal homogenization.
The basic idea is to choose the function ${\hat \varepsilon}_{\parallel}\left(k_x\right)$ so as to enforce the matching between (\ref{eq:hatchi1}) and (\ref{eq:Chi1}) up to the fourth order in $d$. After some algebra, this yields a quadratic equation in ${\hat \varepsilon}_{\parallel}^2$, viz.,
\beq
k^2d^2 {\hat \varepsilon}_{\parallel}^2\left(k_x\right)-2\left(
6+k_x^2d^2
\right){\hat \varepsilon}_{\parallel}\left(k_x\right)+2
\left(
6+k_x^2d^2
\right){\bar \varepsilon}_{\parallel}-k^2d^2 \alpha_a\alpha_b=0,
\eeq
with the coefficients $\alpha_a$ and $\alpha_b$ defined in (\ref{eq:alphas}).
Of the two possible solutions, one turns out to be physically inconsistent, as it diverges in the limit $kd\rightarrow 0$. We are therefore left with
the solution in (\ref{eq:epsNL}), which consistently reduces to the standard (local) EMT model (\ref{eq:EMT}) in the above limit.
This yields an initial (unit-cell) trace error
\begin{eqnarray}
\Delta\chi_1=\chi_1-{\hat \chi}_1
&\approx&
\frac{\left(
	kd\right)^6\left(\varepsilon_a-\varepsilon_b
\right)^2f_a^2f_b^2
}{360}\nonumber\\
&\times&
\left\{
3\varepsilon_a+f_b\left[
\varepsilon_b-5\varepsilon_a+2f_b\left(
\varepsilon_a+\varepsilon_b
\right)
\right]+\left(\frac{k_x}{k}\right)^2\left(
4f_af_b-3
\right)
\right\},
\end{eqnarray}
as compactly indicated in (\ref{eq:hatDchi1}).


%

\newpage

%

%

%

%

%

%

%

%

%

%

\end{document}